# Dark Matter — Cosmic Microwave Background Connection

Krzysztof M. Górski[1]

Universities Space Research Association, NASA/GSFC, Code 685, Greenbelt MD 20771

gorski@stars.gsfc.nasa.gov

## ABSTRACT

A method for inference of the primordial power spectrum from the $COBE$[2] DMR sky maps is discussed. This approach involves a Fourier decomposition of the sky maps in a basis of orthonormal functions on the incompletely sampled sky, a detailed mathematical model of all noise properties, and a likelihood analysis of the data based on an exact probabilistic model of the measurable quantities, valid in the context of Gaussian theories of structure formation. Results from the two year $COBE$-DMR data analysis and the implications for inflationary (flat geometry) models dominated by either dark matter or a cosmological constant are presented. The cross power spectrum of the 53 and 90 GHz DMR sky maps is also discussed.

## 1. INTRODUCTION

The development of inflationary ideas during the 1980s (Blau & Guth 1987) induced a decade-long adherence to the cosmological paradigm which posits that the universe is spatially flat. Such a picture requires that the present energy density of the universe is dominated by non-baryonic dark matter or alternatively by a non-zero vacuum energy contribution (a cosmological constant term, $\Lambda$). The minimal version of the model, which invokes cold dark matter (CDM) as the major constituent of the universe, is presently in direct confrontation with astronomical observations. An extension of the model, which in addition to CDM postulates an admixture of hot dark matter (HDM), enjoys considerable popularity in contemporary cosmological research. Vigorous discussion ensues in the literature as to the plausibility of the mixed dark matter (MDM) model as a viable cosmology, (see e.g. Schaefer, Schafi & Stecker 1989; Davis, Summers, & Schlegel, 1992; Klypin et al. 1993; Pogosyan & Starobinskii 1993; Ma & Bertschinger 1994; Primack et al. 1994; Ma, and Klypin et al. in this volume). Cosmological constant dominated, spatially flat, cold dark matter models (CDM-$\Lambda$ ) can be considered a natural extension of the CDM scenario. By adjusting the combination of a non-zero cosmological constant and the total matter density parameter, $\Omega_0$ (which remains poorly constrained by observations), we can maintain the spatial flatness of the universe required by the inflationary paradigm, whilst striving to improve the agreement of the theoretical model predictions to the observational data (Peebles 1984; Efstathiou, Maddox, & Sutherland 1991; Kofman, Gnedin,

---

[1]On leave from Warsaw University Observatory, Aleje Ujazdowskie 4, 00-478 Warszawa, Poland

[2]The National Aeronautics and Space Administration/Goddard Space Flight Center (NASA/GSFC) is responsible for the design, development, and operation of the Cosmic Background Explorer (COBE). Scientific guidance is provided by the COBE Science Working Group. GSFC is also responsible for the development of the analysis software and for the production of the mission data sets.

Bahcall 1993; Carrol, Press, Turner 1992). The empirical determination of the power spectrum of primordial inhomogeneities and assessment of their consistency with the predictions of these inflation-related models of the universe are critical issues in contemporary cosmology.

The implications of recent observational advancements in extragalactic astronomy, including measures of the galaxy distribution and bulk flow motions for the above models of the universe are discussed elsewhere in this volume. In this contribution I focus on some issues related to the current status of the cosmic microwave background (CMB) anisotropies. White, Scott, & Silk (1994) have recently reviewed both the theoretical and experimental status of the field — the interested reader is invited to refer to this paper. In recent years there has been rapid growth in the experimental area of medium and small angular scale CMB anisotropy, but the field is presently in a transitory stage. Therefore I shall concentrate on the $COBE$ DMR results and their relation to Dark Matter in the universe.

The $COBE$ DMR discovery of CMB anisotropy (Smoot et al. 1992, Bennett et al. 1992, Wright et al. 1992) has affected cosmology in both ontological and practical ways, but its predominant quantitative influence has been to provide the means for the accurate normalisation of theories of large scale structure formation. In the following I will discuss the issues of power spectrum estimation from the $COBE$ DMR sky maps, the related issue of model normalisation and implications for the inflationary models of structure formation.

## 2. CMB ANISOTROPY POWER SPECTRUM ESTIMATION

### 2.1 $COBE$ DMR Data

All the results discussed were derived by analysis of the $COBE$ DMR two year sky maps. A summary of two years of observations has been presented in Bennett *et al.* (1994). The $COBE$ DMR instrument has observed the microwave sky at three frequencies (31.5, 53, and 90 GHz) with pairs of $\sim 7°$ (FWHM) antennas separated by 60° on the sky. The temperature difference measurements were used to construct whole sky anisotropy maps (Janssen & Gulkis 1992) binned into 6144 equal area ($\sim 2°\!.6 \times 2°\!.6$) pixels. The 31 GHz maps were not used in the analysis, since they are relatively noisy and contain a more significant galactic signal contribution. The $A$ and $B$ channels of the sky maps are coadded using inverse noise variance weighting. The sky maps have been generated in both galactic and ecliptic coordinate frames, involving separate binning of the underlying time-ordered series of sky temperature measurements into 6144 sky map pixels (of equal area). In this analysis, both the galactic and ecliptic coordinate frame data sets are used as a check on the extent to which the coordinate dependent noise binning can affect the inferred normalisation. Strong galactic plane emission was excised from the maps by removing all pixels within $|b| = 20°$ of the galactic equator. Faint, high latitude galactic emission was not removed from the data (since no fully supported models or definitive measurements thereof exist). As a simple measure of the extent to which such foreground emission could affect the inferred normalisation, the power spectrum analysis is conducted with the lowest order ($\ell = 2$) anisotropy mode, the quadrupole, either included or excluded. No smoothing or filtering of the sky maps is applied prior to the implementation of the power spectrum inference.

## 2.2 Model Power Spectra of CMB Anisotropy

The cosmological models are specified to within an arbitrary amplitude of the perturbations as follows:
1) the global geometry is flat with $\Omega_0 + \Omega_b + \Omega_\Lambda = 1$, with the bulk mass density provided by either CDM or MDM; for the MDM models (with only $\Lambda = 0$ considered) the hot dark matter is introduced in the form of either one or two (equal mass) families of massive neutrinos, with the contributed fraction of critical density taken as $\Omega_\nu = 0.15$, 0.2, 0.25 and 0.3 for one massive flavour, and $\Omega_\nu = 0.2$, 0.3, otherwise;
2) the value of the Hubble constant, $H_0 = 100\,h$ km s$^{-1}$ Mpc$^{-1}$, is sampled at $h = 0.5$, and 0.8, and, in agreement with the Big-Bang nucleosynthesis arguments, the baryon abundance obeys the relation $\Omega_b = 0.013\,h^{-2}$ (Reeves 1994);
3) random-phase, Gaussian, scalar primordial curvature perturbations (no gravity waves) are assumed with the inflationary Harrison-Zel'dovich spectrum corresponding to an adiabatic density perturbation distribution, $P(k) \propto k$.

The angular distribution of CMB temperature anisotropy induced by such curvature perturbations (Sachs & Wolfe 1967, SW) is easily described in Fourier language, using the spherical harmonic decomposition of the random field of temperature fluctuations — $\delta T/T(\theta, \phi) = \sum_{\ell m} a_{\ell m} Y_{\ell m}(\theta, \phi)$ (e.g. Peebles 1981). Individual spherical harmonic coefficients $a_{\ell m}$ are Gaussian-distributed in the theoretical ensemble of initial conditions for structure formation. The variances of the probability distributions of individual modes, $\langle |a_{\ell m}|^2 \rangle$, are uniquely expressed as integrals over the power spectrum, and depend only on $\ell$ due to the statistical isotropy of the CMB temperature field. The CMB anisotropy multipole coefficients and the matter perturbation transfer functions for all models were evaluated using the Boltzman equation code of Stompor (1994) by solving the propagation equations up until the redshift $z = 0$.

Over the low-$\ell$ range of CMB multipoles probed by $COBE$-DMR the theoretical spectra are indistinguishable for the CDM and MDM models with equivalent $h$ and $\Omega_b$. Thus, the power spectrum amplitude derived from the data applies equally to both CDM and MDM models. The low-$\ell$ shape of the CDM/MDM spectrum is a little steeper than the underlying, inflationary $n = 1$ spectrum (Bond 1993; Bunn, Scott, & White 1994; Górski, Stompor, & Banday 1995). The steepening is caused by three effects: 1) the anisotropy of the photon distribution function at the moment of decoupling, 2) the high-redshift integrated SW effect generated by the change in the growth rate of matter perturbations at the post-recombination epoch due to the non-negligible contribution of radiation (i.e. photons and massless neutrinos) to the total energy density, 3) the contribution of the first adiabatic/Doppler peak to the low $\ell$ multipoles. These contributions, though small compared to the overall amplitude of the usual curvature perturbation driven SW effect, should be accounted for in an accurate power spectrum normalisation based on the $COBE$-DMR anisotropy data. The CDM-$\Lambda$ radiation power spectra at low-$\ell$ ($\lesssim 10$), with a characteristic enhancement of the lowest order modes, are determined by both the usual SW and the cosmological constant induced, integrated SW (ISW) effects (Kofman & Starobinskii 1985, Górski, Silk, & Vittorio 1992), and depend very weakly on the Hubble constant.

We express the power spectrum amplitude in terms of $Q_{rms-PS}$, the value of the exactly computed quadrupole, $a_2$, multiplied by $(5/4\pi)^{1/2}$, which depends non-trivially on cosmological parameters. This is a generalisation of the normalisation introduced in Smoot et al. (1992) for pure power law model spectra. For

$\Omega_0 \lesssim 0.5$, the ISW contributions boost the quadrupole over its pure SW value. For $\Omega_0 \gtrsim 0.5$ — models with a negligible cosmological constant induced ISW effect — the resulting exact quadrupole is lower than its pure SW counterpart. The steepest spectra arise in models near critical matter density ($\Omega_0 \simeq 0.8$).

2.3 Power Spectrum Parameter Inference

The power spectrum estimation method described in Górski (1994) was applied. Orthogonal basis functions are constructed from spherical harmonics with the exact inclusion of both pixelisation effects and galactic plane excision, which is coordinate system specific, and leaves 4016 and 4038 pixels in the galactic and ecliptic sky maps, respectively. [Note that unlike the Karhunen-Loeve, or signal-to-noise eigenmode approach (Bond 1993, Bunn and Sugiyama 1994) this approach utilizes mathematical constructs which are *entirely* independent of the properties of the theory which one wishes to test with the data.] The resulting finite Fourier series decomposition of the signal over the spectral range $\ell \in [2, 30]$ is then performed in this basis, and the Fourier mode amplitudes of the CMB anisotropy are used in the construction of an exact Gaussian likelihood function for a given theory. As explained in Górski (1994), this method is unique in allowing the algebraic elimination of the non-cosmological, $\ell = 0, 1$, modes (without affecting higher-$\ell$ modes at all) from the power spectrum inference procedure.

The correlation matrices for both theoretical signal and map noise are evaluated using the orthogonal mode expansion, i.e. explicitly reflecting the mode-coupling due to the galaxy cut. The noise is considered to be sufficiently described by a pixel-to-pixel uncorrelated random process (the 60° correlations have a negligible effect on the inference from the DMR data, see Lineweaver et al. 1994).

Monte Carlo testing has demonstrated that this method of power spectrum inference is statistically unbiased (Fig. 1).

2.4 Results of $Q_{rms-PS}$ Fitting and Large Scale Structure Statistics

The actual two parameter contour plot of the likelihood derived in the pure power law power spectrum fit to the two year data is shown in Górski et al. (1994), and in Bennett et al. in this volume. A detailed presentation of the results for CDM, MDM, and CDM-$\Lambda$ model fitting to the $COBE$ DMR data can be found in Górski, Stompor, & Banday (1995), and Stompor, Górski, & Banday (1995). For comparison one may refer to Bunn, Scott, & White(1994), and Bunn & Sugiyama (1994). A typical one parameter likelihood fit of a flat, CDM/MDM model to the two year $COBE$-DMR data yields a $\sim 13\sigma$ significant determination of $Q_{rms-PS} \sim 20\,\mu$K. For the CDM-$\Lambda$ models the best fit amplitude grows with decreasing $\Omega_0$ up to $\sim 26\,\mu$K. Systematic shifts in the central value of the fit are observed due to: 1) differences in the noise pixelisation in the galactic and ecliptic coordinate frames, which result in a $\sim 0.8\,\mu$K difference between the inferred normalisation amplitudes, with higher values obtained from the ecliptic maps; 2) exclusion of the quadrupole, which produces a $\sim +0.4\,\mu$K variation in the fitted amplitude; 3) the uncertainty in the values of $h$ and $\Omega_b$ (reflected by differences in spectral shape over the $\ell$-range accessible to DMR) causes an additional small spread, $\pm 0.25\,\mu$K, of the fitted amplitudes.

A convenient summation of the proposed overall normalisation for the CDM and MDM models is $Q_{rms-PS} = (20 \pm 1.52)(\pm 0.4 \pm 0.2 \pm 0.25)\mu$K, and for the CDM-$\Lambda$ models $Q_{rms-PS} \simeq \{[20 + 9.1h^{0.6}\exp(-\Omega_0^2 h^{0.3}/0.057)] \times (1 \pm 0.077) \pm 0.4 \pm 0.2 \pm 0.25\}\mu$K. The error ranges represent the statistical error and uncertainties associated with effects 1 through 3 above, respectively. It will be noted that the statistical error on the inferred normalisation is considerably larger than the other uncertainties.

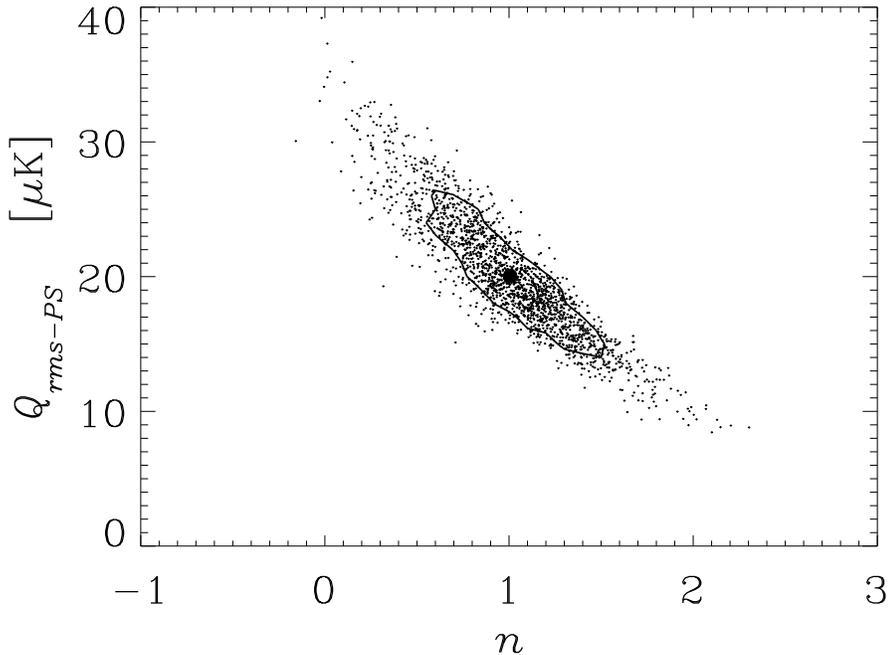

Figure 1: Results from Monte Carlo simulations of the power spectrum inference procedure. Each of the 2000 points shows the locus of the maximum likelihood parameters derived from a simulated CMB sky map for an input model with $n = 1$, and $Q_{rms-PS} = 20\,\mu K$ and noise appropriate to the DMR 53 GHz receivers. The filled circle represents the average of the output parameters and demonstrates that the inference method is statistically unbiased. The solid line is the approximate 68% probability density contour derived from the simulated distribution.

Table 1 contains a representative selection of models and the rms values of several large-scale structure statistics computed from the matter perturbation spectra according to the usual prescriptions given in the footnotes.

The cold dark matter theory with a standard choice of cosmological parameters requires a very high normalisation in order to fit the CMB anisotropy distribution. Analysis of the first year of *COBE*-DMR data had already suggested that $(\sigma_8)_{mass} \sim 1$ (Wright et al. 1992, Efstathiou, Bond, & White 1992), and this value increases to $\sim 1.4$ with two years of data and an improved analysis technique. Although this normalisation allows the theory to predict POTENT-scale velocities (Dekel 1994) with a comfortably high amplitude, it is still not high enough to explain the Lauer & Postman (1994) result, and it also results in a significant overproduction of density perturbations on scales of $\lesssim 20\,h^{-1}\,\mathrm{Mpc}$. CDM has often been criticised for its poor match to both galaxy and cluster distributions. Mixed dark matter models manage to circumvent, to a certain degree, these same problems by construction — massive neutrinos partially damp the perturbations at those length-scales where CDM looks problematic. Among the MDM models those with two species of massive neutrino seem to meet the observational constraints more comfortably (see also Primack et al. 1994). The larger free-streaming radius allows for the suppression of the perturbation amplitude on larger scales than in models with one massive flavour. This is reflected in the decrease of the predicted $(\sigma_8)_{mass}$ values. Nevertheless, the proponents of Dark Matter cosmology will have to address the viability of the models viz. the contradictory requirements that there *should be* a bias between the galaxy and mass distribution, as suggested by the galaxy cluster abundance argument (White, Efstathiou, & Frenk 1993) and the galaxy pair-wise velocities, and there *should not be* one, as suggested by the $COBE$ DMR normalisation.

The low-$\Omega_0$ CDM-$\Lambda$ models can not be rejected solely on the basis of the $COBE$-DMR data ($\Omega_0 \geq 0.15$, 95% confidence), although the most likely $\Omega_0$ value is strikingly close to unity. Conversely, observations of the matter distribution do require low values of the total density, with $\Omega_0 h^2 \sim 0.1$. being the favoured value. On larger scales, these models predict bulk flows in reasonable agreement with POTENT, but dramatically smaller than determined by the Lauer & Postman analysis. Therefore, if the latter observation is confirmed all CDM-$\Lambda$ models (together with critical density CDM or MDM models) will be found wanting.

At present, although the position of the CDM-$\Lambda$ models is more comfortable than that of critical density models, it is not free from potentially fatal flaws. Unfortunately, CMB anisotropies on a one degree scale do not offer any serious prospects for distinguishing between cosmological constant models and other viable scenarios (Stompor & Górski, 1994; Bond et al., 1994). Hence, definite conclusions will have to await more reliable observational data, particularly large-scale, deep galaxy surveys capable of unraveling the shape of the galaxy power spectrum down to $k \sim 0.01\, h\text{Mpc}^{-1}$, or a comoving length-scales up to $\sim 600\, h^{-1}\text{Mpc}$.

## 3. CROSS POWER SPECTRUM

Visualising the faint CMB structure present in the rather noisy DMR maps is a difficult task. A straightforward and relatively simple procedure is the computation of a 'cross' power spectrum between two sky maps. Such a construction can take full advantage of the multiple channel data acquisition by the DMR instrument. Since the noise properties of the different channels are not correlated, there is no noise bias in the amplitude of the quadratic power spectrum coefficients.

The cross power spectrum presented here is derived by the direct product of the individual Fourier coefficients from the 53 and 90 GHz sky maps. Monte Carlo simulations are performed to obtain comparison statistical distributions for the CDM (MDM) model, since analytic solutions are no longer easily computed.

Figure 2 displays both the sum (A+B) and difference (A-B) map cross power spectra, and clearly demonstrates a significant detection of CMB anisotropy by the $COBE$-DMR instrument. The signal-to-noise ratio in the cross-combined 53 and 90 GHz two year data exceeds unity for $\ell \lesssim 15$. The cross spectra derived in galactic and ecliptic coordinates appear to be in excellent agreement. This is not a trivial fact — the basis functions with respect to which the Fourier decomposition was performed are coordinate frame dependent due to the different geometry of the applied Galactic cut. Thus Fig. 2 is the best demonstration to date of the rotational invariance of the angular spectrum of *COBE* DMR measured CMB anisotropy.

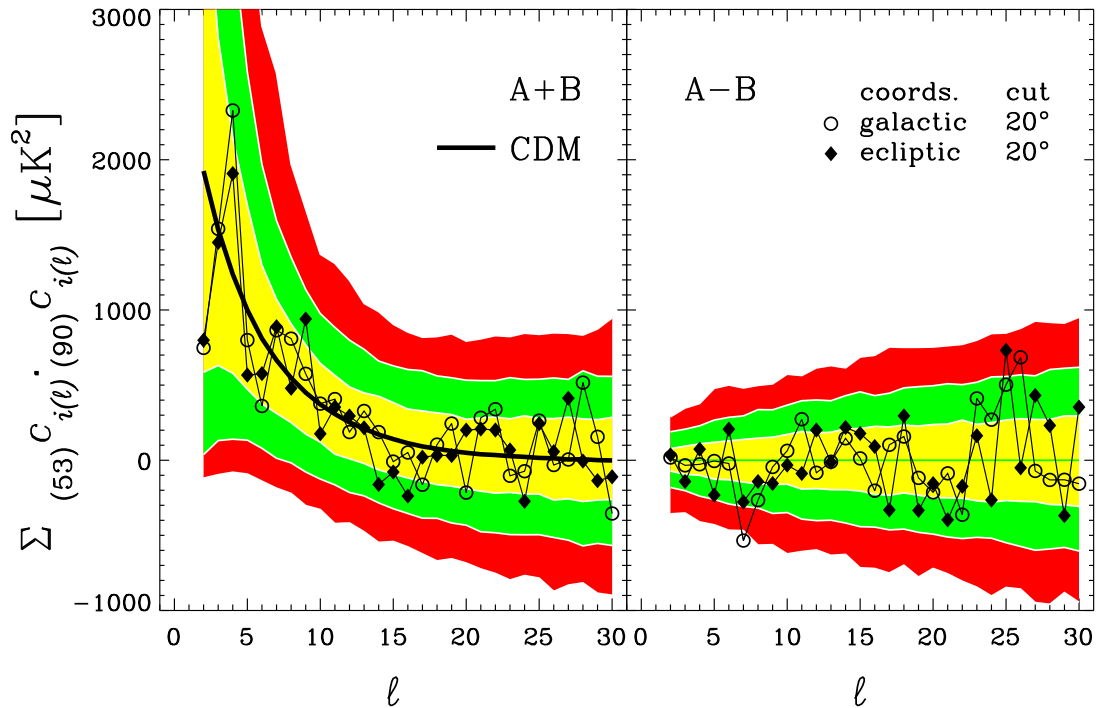

Figure 2: The cross power spectra for the DMR 53 and 90 GHz two year sum and difference sky maps derived in both galactic and ecliptic coordinate frames. The results of 40,000 Monte Carlo simulations of CDM skies normalised to $Q_{rms-PS} = 19.5\,\mu K$ and with noise appropriate to the two year sky maps are shown. The heavy, solid line shows the mode of the power distribution at each $\ell$. Note that this is *not* a fit to the plotted cross-power spectrum. The grey bands represent the confidence regions for the distribution of power at each value of $\ell$ as follows: light grey — 68%, medium — 95%, and dark grey — 99.8%.

The overplotted CDM power spectrum distribution (identical to MDM over this range of $\ell$) appears to be a reasonable statistical descriptor for the measured CMB anisotropy pattern. The low *COBE* DMR two year quadrupole amplitude (Bennett et al. 1994) is clearly evident, despite the fact that we have not attempted to model and correct for residual high latitude Galactic emission. Although such modelling would result in a lower estimated cosmological CMB quadrupole amplitude, it would still fall within the 95% probability zone under the distribution derived from the *COBE* DMR normalised CDM/MDM model. An analogous plot for CDM-$\Lambda$ models would only show differences at low $\ell$ for low values of $\Omega_0$; the average power in this range would tend to be higher than in Fig. 2. Coupled with the relatively low measured value of the quadrupole, the relative likelihood of high-$\Lambda$ models is lower than that for the matter dominated models.

## 5. SUMMARY

The improved quality of the two year $COBE$-DMR data combined with reliable power spectrum estimation techniques allows the accurate normalisation of cosmological theories. Previously, the variance of $COBE$-DMR temperature fluctuations on a 10° angular scale was utilised for the normalisation of the power spectrum. Subsequent work (Wright et al. 1994, Banday et al. 1994, and these proceedings) has demonstrated that this technique can be unreliable without considerable attention. More appropriate methods take full advantage of the measured CMB anisotropy power distribution on all angular scales accessible to the $COBE$-DMR instrument, as discussed in this contribution.

Whilst it would be safer to await the final 4-year COBE results before offering definitive statements as to the viability of theoretical models, one should note that the CDM normalisation derived from the two year $COBE$-DMR data does appear to be irreconcilably high, while the MDM and CDM-$\Lambda$ models have little room left for adjustments. Interestingly, theoretically favoured globally flat universe models notwithstanding, it is possible to construct open universe inflationary dark matter models in reasonable agreement with the observations (Górski, Ratra, Sugiyama, & Banday 1995, and references therein).

## ACKNOWLEDGEMENTS

I am grateful to R. Stompor and A.J. Banday for allowing me to use the results from our joint work before publication. I am grateful to A.J. Banday, C. Bennett, G. Hinshaw and A. Kogut for numerous discussions and help in improving the manuscript. The efforts of those contributing to the $COBE$-DMR are acknowledged.

Table 1: Inferred cosmological statistics for spatially flat CDM, MDM and CDM$\Lambda$ models (with $\Omega_0 + \Omega_\Lambda + \Omega_b = 1$, where $\Omega_0 = \Omega_{CDM} + \Omega_\nu$ and $\Omega_b = 0.013 h^{-2}$), with a $COBE$-DMR normalisation expressed in terms of $Q_{rms-PS}$. $N_\nu$ is the number of massive neutrino species, and $m_\nu$ the neutrino mass in eV. The errors, including both statistical ($1\sigma$) and systematic deviations, are of the order of 11%.

| $\Omega_0$ | $h$ | $\Omega_\nu$ | $N_\nu$ | $m_\nu$ | $Q^{(a)}_{rms-PS}$ | $(\sigma_8)^{(b)}_{mass}$ | $J_3(20)^{(c)}$ | $v^{(d)}_{40}$ | $v^{(d)}_{60}$ | $v^{(d)}_{100}$ |
|---|---|---|---|---|---|---|---|---|---|---|
| 1.0 | 0.5 | 0.00 | – | 0.0 | 20.04 | 1.36 | 968 | 444 | 355 | 248 |
| .. | .. | 0.15 | 1 | 3.7 | .. | 0.97 | 706 | 441 | 357 | 251 |
| .. | .. | 0.20 | 1 | 4.9 | .. | 0.92 | 694 | 442 | 358 | 252 |
| .. | .. | 0.25 | 1 | 6.1 | .. | 0.88 | 691 | 444 | 359 | 252 |
| .. | .. | 0.30 | 1 | 7.3 | .. | 0.85 | 695 | 445 | 360 | 252 |
| .. | .. | 0.20 | 2 | 2.4 | .. | 0.82 | 567 | 435 | 356 | 252 |
| .. | .. | 0.30 | 2 | 3.7 | .. | 0.71 | 516 | 439 | 359 | 254 |
| 0.1 | 0.8 | 0.0 | – | 0.0 | 26.11 | 0.53 | 335 | 208 | 192 | 167 |
| .. | 0.5 | .. | .. | .. | 25.27 | 0.17 | 41 | 139 | 134 | 123 |
| 0.2 | 0.8 | .. | .. | .. | 23.61 | 1.03 | 985 | 330 | 290 | 230 |
| .. | 0.5 | .. | .. | .. | 23.24 | 0.44 | 222 | 235 | 216 | 183 |
| 0.3 | 0.8 | .. | .. | .. | 22.04 | 1.39 | 1496 | 403 | 342 | 258 |
| .. | 0.5 | .. | .. | .. | 21.79 | 0.66 | 428 | 301 | 267 | 214 |
| 0.4 | 0.8 | .. | .. | .. | 21.10 | 1.66 | 1825 | 449 | 372 | 272 |
| .. | 0.5 | .. | .. | .. | 20.93 | 0.84 | 607 | 349 | 301 | 233 |

(a) in $\mu K$

(b) $(\sigma_{hR})^2_{mass} = \frac{1}{2\pi^2} \int_0^\infty w^2_{TH}(kR) P(k) k^2 \, dk$

(c) $J_3(hR) = \frac{R^3}{2\pi^2} \int_0^\infty w^2_{TH}(kR) P(k) k^2 \, dk$, in $(h^{-1}\mathrm{Mpc})^3$, and for $R = 20 h^{-1}\mathrm{Mpc}$

(d) $v^2_{hR} = \frac{H_0^2}{2\pi^2} \int_0^\infty w^2_{TH}(kR) e^{-k^2 r_s^2} P(k) \, dk$, $hr_s = 12\mathrm{Mpc}$, in km s$^{-1}$

$w_{TH}(x) = 3 j_1(x)/x$

# COBE[1]-DMR Two-Year Large Scale Anisotropy Results

C.L. Bennett, A. Banday, K. Górski, G. Hinshaw, A. Kogut
Code 685, Laboratory for Astronomy and Solar Physics
NASA/Goddard Space Flight Center, Greenbelt MD 20771
bennett@stars.gsfc.nasa.gov

& E.L. Wright
Department of Astronomy
University of California, Los Angeles, CA 90024

ABSTRACT

We summarize the results of the Cosmic Microwave Background (CMB) anisotropy measurements of the COBE-DMR using the first two years of data. The inference of angular power spectrum parameters from the DMR data is complicated by the need to omit Galaxy contaminated regions of the sky, preventing the spherical harmonics from forming an orthogonal basis set over the remaining sky. The COBE-DMR team has implemented three methods for deducing the power spectrum parameters. These include: a two-point correlation function analysis (Bennett et al. 1994); an analysis in terms of orthogonal modes constructed on the cut sphere (Górski et al. 1994); and an analysis of a modified Hauser-Peebles power spectrum (Wright et al. 1994). These methods for deducing the power spectrum are discussed and the results are summarized and compared.

## 1. INTRODUCTION

The large angular scale cosmic temperature fluctuations are believed to reflect the gravitational potential fluctuations at the epoch of recombination. The amplitude, $Q_{rms-PS}$, and power law spectral index, $n$, of the primordial anisotropy are important cosmological parameters, and serve as initial conditions for for N-body and hydrodynamic codes that test dark matter models. We have applied three methods for determining the amplitude and slope of the primordial power spectrum to the two-year COBE-DMR data. The results from each method are presented in detail in Table 1. The three methods yield maximum likelihood values for $n$ between 1.02 and 1.42 with uncertainties between 0.32 and 0.45. In no case is a determination inconsistent with the scale-invariant value $n = 1$, and the steeper determinations tend to be driven by the smallness of the quadrupole. All methods return a similar ridge of degeneracy between $Q_{rms-PS}$ and $n$, though the slope of the ridge does vary slightly from method to method. The maximum likelihood values for the normalization of an $n = 1$ spectrum range from 18.2 to 20.4 $\mu$K, with uncertainties between 1.6 and 2.0 $\mu$K. The $n$ values returned by each method vary by about 0.5 $\sigma$, while the normalizations for a scale-invariant spectrum vary by about 1 $\sigma$.

The two-point correlation analysis and the modified Hauser-Peebles power spectrum analysis employ maximum likelihood fits to quadratic statistics constructed from products of maps: a combination of the 53 and 90 GHz data. The

---

Table 1: Power Spectrum Analysis of Combined 53 and 90 GHz 2-yr DMR Data

| Parameter | 2-pt function Bennett et al. | $\psi_{\ell m}$ modes Górski et al. | $G_{\ell m}$ modes Wright et al. |
|---|---|---|---|
| | including quadrupole | | |
| $n$ | $1.42^{+0.49}_{-0.55}$ | $1.22^{+0.43}_{-0.52}$ | $---$ |
| $n$ (marginal[a]) | $1.42 \pm 0.37$ | $1.10 \pm 0.32$ | $1.39^{+0.34}_{-0.39}$ |
| $Q_{rms-PS}$ ($\mu$K) | $14.3^{+5.2}_{-3.3}$ | $17.0^{+7.6}_{-4.8}$ | $---$ |
| $Q_{rms-PS\|n=1}$ ($\mu$K) | $18.2 \pm 1.5$ | $19.9 \pm 1.6$ | $---$ |
| | excluding quadrupole | | |
| $n$ | $1.11^{+0.60}_{-0.55}$ | $1.02^{+0.53}_{-0.59}$ | $---$ |
| $n$ (marginal) | $1.11 \pm 0.40$ | $0.87 \pm 0.36$ | $1.25^{+0.40}_{-0.45}$ |
| $Q_{rms-PS}$ ($\mu$K) | $17.4^{+7.5}_{-5.2}$ | $20.0^{+10.5}_{-6.5}$ | $---$ |
| $Q_{rms-PS\|n=1}$ ($\mu$K) | $18.6 \pm 1.6$ | $20.4 \pm 1.7$ | $19.8 \pm 2.0$ |
| $a_\ell$ at pivot ($\mu$K)[b] | $a_7 = 9.5 \pm 1.0$ | $a_9 = 8.3 \pm 0.7$ | $---$ |

[a]Bennett et al. and Wright et al. define the marginal likelihood on $n$ to be $L(n) = \max L(Q|n)$, while Górski et al. define it to be $L(n) = \int_0^\infty L(Q,n) dQ$, where $Q \equiv Q_{rms-PS}$.

[b]An alternate power spectrum normalization, expressed in terms of the multipole moment that is independent of $n$.

techniques make extensive use of Monte Carlo simulations to construct and calibrate the derived likelihood functions. The orthogonal mode analysis employs an exact maximum likelihood fit to the linear coefficients $c_i$ from the 53 and 90 GHz data, which, in effect, places the signal and noise on a similar footing. Monte Carlo simulations are only used to verify that the technique is statistically unbiased. Note that the three methods will tend to respond differently to unmodeled effects in the data, such as residual Galactic emission, or systematic artifacts, though we believe that any such effects are small. While the three methods return generally consistent results, small discrepancies do exist and the reasons for this are not presently understood. Studies of our methodologies are continuing as we prepare to analyze the full four-year sky maps.

## 2. TWO-POINT CORRELATION FUNCTION

The two-point cross correlation function at angular separation $\alpha$, $C(\alpha)$, is the average product of all pixel temperatures with a fixed angular separation: $C(\alpha) = \sum_{i,j} w_i w_j T_{A,i} T_{B,j} / \sum_{i,j} w_i w_j$, where the sum is restricted to pixel pairs $(i,j)$ for which the pixel angular separation resides in the bin defined by $\alpha$, $T_{A,i}$ and $T_{B,i}$ are the observed temperature in maps "A" and "B" respectively, after monopole and dipole (and quadrupole) subtraction, and $w_i$ is the statistical weight of pixel $i$. The analysis discussed below is based on the 53 × 90 GHz cross correlation function.

We determine the most likely quadrupole normalized amplitude, $Q_{rms-PS}$, and spectral index, $n$, by evaluating the Gaussian approximation to the likelihood

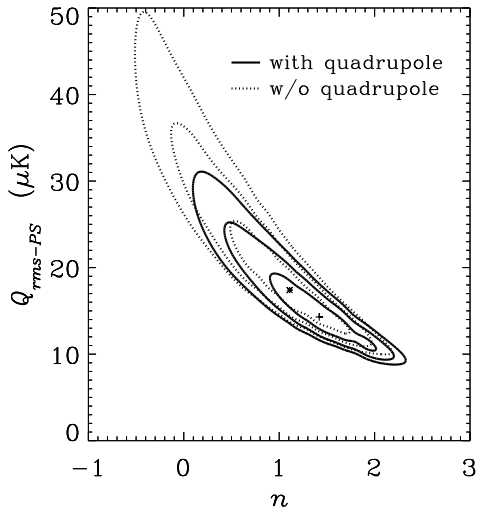

Figure 1: Likelihood contours as a function of $n$ and $Q_{rms-PS}$ for the $53(A+B)/2 \times 90(A+B)/2$ cross correlation data. The contours correspond to 68%, 95%, and 99.7% confidence regions, corrected for a bias in $n$ of $+0.11$, and in $Q_{rms-PS}$ of $-1.5$ $\mu$K. The solid curves are contours with the quadrupole included in the analysis; the peak of the likelihood is indicated by $+$. The dashed curves are contours with the quadrupole excluded from the analysis; the peak of the likelihood is indicated by $*$. See Table 1 for a quantitative summary of these likelihoods.

function

$$L(Q_{rms-PS}, n) = \frac{\exp[-\frac{1}{2}(\boldsymbol{\Delta C})^T \mathbf{M}^{-1}(\boldsymbol{\Delta C})]}{[(2\pi)^m \det(\mathbf{M})]^{1/2}}.$$

Here $(\boldsymbol{\Delta C})^T$ and $(\boldsymbol{\Delta C})$ are row and column vectors with entries $(\Delta C)_i = C(\alpha_i) - \langle C(\alpha_i) \rangle$, and $\mathbf{M} = \langle (\boldsymbol{\Delta C})(\boldsymbol{\Delta C})^T \rangle$ is the covariance matrix of the correlation function. The angled brackets denote averages over receiver noise and cosmic variance. We estimate the mean correlation function and covariance matrix, as a function of $Q_{rms-PS}$ and $n$, by means of Monte Carlo simulations that account for all important aspects of our data processing. The final results are calibrated with additional Monte Carlo simulations to account for any biases introduced by the Gaussian approximation or other effects. Figure 1 (after Figure 4 of Bennett et al.) shows the resulting likelihood contours as a function of $Q_{rms-PS}$ and $n$ for the analyses both including and excluding the quadrupole moment. The contours correspond to 68%, 95%, and 99.7% confidence regions.

## 3. ORTHOGONAL MODE DECOMPOSITION

Górski et al. explicitly construct linear combinations of spherical harmonics to form orthonormal functions on the cut sphere, following a procedure developed by Górski (1994). The 53 and 90 GHz data are decomposed with respect to these modes: $T(p) = \sum_i c_i \psi_i(p)$, where $T(p)$ is the temperature in map pixel $p$, and $c_i$ is the coefficient of the $ith$ mode $\psi_i$. For Gaussian power-law models of anisotropy, the $c_i$ coefficients have a multivariate Gaussian distribution, thus an exact likelihood function for the parameters $Q_{rms-PS}$ and $n$ may be constructed

$$L(Q_{rms-PS}, n) = \frac{\exp[-\frac{1}{2}\hat{\mathbf{c}}^T \cdot (\mathbf{C_{CMB}} + \mathbf{C_N})^{-1} \cdot \hat{\mathbf{c}}]}{[(2\pi)^m \det(\mathbf{C_{CMB}} + \mathbf{C_N})]^{1/2}}$$

where $\hat{\mathbf{c}}$ is the vector of $c_i$ coefficients obtained from decomposing the 53 and 90 GHz data, and $(\mathbf{C_{CMB}} + \mathbf{C_N})$ is the covariance matrix of the spectra $\hat{\mathbf{c}}$, which depends on $Q_{rms-PS}$ and $n$. See Górski and Górski et al. for details. Monte Carlo simulations with controlled inputs indicate that the parameters inferred with the above likelihood function are not statistically biased with respect to the input. Figure 2 (after Figure 2 of Górski et al.) shows the resulting likelihood contours as a function of $Q_{rms-PS}$ and $n$ for the analyses both including and excluding the quadrupole moment. The contours correspond to 68%, 95%, and 99.7% confidence regions.

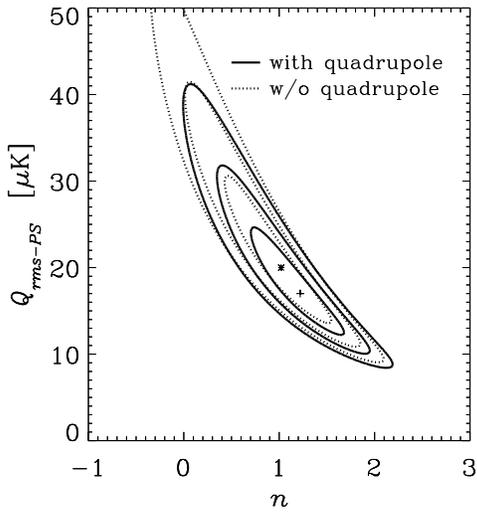

Figure 2: Likelihood contours as a function of $n$ and $Q_{rms-PS}$ for the 53 and 90 GHz orthogonal mode decomposition. The contours correspond to 68%, 95%, and 99.7% confidence regions. The solid curves are contours with the quadrupole included in the analysis; the peak of the likelihood is indicated by $+$. The dashed curves are contours with the quadrupole excluded from the analysis; the peak of the likelihood is indicated by $*$. See Table 1 for a quantitative summary of these likelihoods.

### 4. MODIFIED HAUSER-PEEBLES POWER SPECTRUM

Wright et al. solve for the angular power spectrum of the DMR two year data by modifying and applying the technique described by Peebles (1973) and Hauser & Peebles (1973) for data on a cut sphere. The input data for this analysis is a "cross" power spectrum which involves the evaluation of terms of the form $\langle G_{\ell m} T_A \rangle \cdot \langle G_{\ell m} T_B \rangle$ where $T_A$ and $T_B$ are two independent DMR maps, $G_{\ell m}$ is a modified spherical harmonic which, by construction, is orthogonal to the monopole and dipole (and, optionally, the quadrupole) on the cut sky, and the angle brackets denote an inner product on the cut sky. See Wright et al. for more details. They have evaluated a Gaussian likelihood, similar to eq. 1, over a variety of spectral ranges: $3 \leq \ell \leq 19$, $3 \leq \ell \leq 30$, and $2 \leq \ell \leq 30$, and for a variety of data combinations. After calibrating their results with Monte Carlo simulations they quote most likely spectral indices in the range 1.02 to 1.41 depending on the choice of data, the spectral range included, and the type of modes used. Their results, together with the other two methods are summarized in Table 1.

# ON THE RMS ANISOTROPY AT 7° AND 10° OBSERVED IN THE $COBE$-DMR TWO YEAR SKY MAPS

A.J. Banday, K.M. Górski, A. Kogut, G. Hinshaw, and C.L. Bennett
Code 685, NASA Goddard Space Flight Center, Greenbelt, MD 20771
C.H. Lineweaver, G.F. Smoot and L. Tenorio
LBL, SSL & CfPA, Bldg 50 - 351, University of California, Berkeley CA 94720

## ABSTRACT

We summarize the recent results on the observed $COBE$-DMR two year sky rms temperature fluctuations. A 'cross-RMS' statistic is used to infer the $Q_{rms-PS}$ normalization for a scale-invariant ($n = 1$) spectral model. The method is extended to the normalization of other cosmological power spectra.

## 1. INTRODUCTION

In principle, the observed sky rms on a given angular scale is a convenient number to use for the normalization of a particular cosmological model. Wright et al. (1994a) used the sky rms temperature fluctuations from the first year $COBE$-DMR sky maps smoothed to an approximate FWHM of 10° to determine the effective normalization $Q_{rms-PS}$ for the scale-invariant Harrison-Zel'dovich power spectrum, $P(k) \propto k^n$, where $n = 1$ and $k$ is the comoving wavenumber. We have updated and extended the analysis of Wright et al. (1994a) using the two year $COBE$-DMR data to infer the normalization for an $n = 1$ power law model, together with a number of other cosmological power spectra.

## 2. ANALYSIS

Banday et al. (1994) have defined a cross-RMS, $\otimes_{RMS}$, between two maps $a$ and $b$ as

$$(\otimes_{RMS})^2 \equiv \sum_i T_i^a T_i^b \, w_i^a \, w_i^b \, / \, \sum_i w_i^a \, w_i^b, \qquad (1)$$

where the sums are over all pixels $i$ surviving the galactic cut ($|b| > 20°$), and $w_i$ is the weight assigned to that pixel. Unit weighting is adopted here. The best fit monopole, dipole and, where appropriate, quadrupole are removed from the sky maps prior to the computation of the $\otimes_{RMS}$.

A Monte Carlo approach was adopted to generate the $\otimes_{RMS}$ distributions for a grid of $Q_{rms-PS}$ values (with 2500 simulations used for each value of $Q_{rms-PS}$). Each simulation combines a realization of sky anisotropy filtered through the $COBE$-DMR beam (Wright et al. 1994a) with noise realizations based on the specific $COBE$-DMR channels. The $Q_{rms-PS}$-dependent statistical means, variances and covariances of the 7° and 10° $\otimes_{RMS}$ are derived from these Monte Carlo simulations, and used to construct the gaussian approximation to the probability distribution of the $\otimes_{RMS}$. This, together with the measured $\otimes_{RMS}$ values, defines the likelihood function $\mathcal{L}(Q_{rms-PS})$. That the $\otimes_{RMS}$ provides a statistically unbiased estimator of the true cosmological $Q_{rms-PS}$ amplitude has been verifed with additional simulations used as test input data.

We refer the reader to Banday et al. (1994) for full details. Summarizing the main results therein:

- The observed $\otimes_{RMS}$ is frequency independent and thus consistent with a cosmic origin for the temperature fluctuations.

- The $\otimes_{RMS}$ values derived from different combinations of the 53 and 90 GHz data are generally in excellent statistical agreement (Table 1), thus the inferred $Q_{rms-PS}$ values are reasonably independent of data selection.

- $Q_{rms-PS}$ derived from the $\otimes_{RMS}$ at 7° and 10° *including* the quadrupole is biased **low** by the small observed $COBE$-DMR sky quadrupole amplitude. This can be corrected for, after which the estimates for $Q_{rms-PS}$ both including and excluding the quadrupole are in excellent agreement.

- the amplitude of primordial inhomogeneity in the context of a Harrison-Zel'dovich $n = 1$ power law spectral model is $Q_{rms-PS} \sim 19$ $\mu$K (in good agreement with values derived from other techniques — see Górski et al. 1994, Bennett et al. 1994, Wright et al. 1994b).

Table 1: Observed $\otimes_{RMS}$ values derived from possible 53 and 90 GHz combinations.

| $\otimes_{RMS}$ combination | including Quadrupole | | excluding Quadrupole | |
|---|---|---|---|---|
| | 7° ($\mu$K) | 10° ($\mu$K) | 7° ($\mu$K) | 10° ($\mu$K) |
| 53A$\otimes$53 | $44.5^{+4.8}_{-4.7}$ | $32.4^{+1.8}_{-1.8}$ | $43.1^{+5.4}_{-5.2}$ | $30.6^{+1.9}_{-1.9}$ |
| 90A$\otimes$90B | $0.0^{+12.0}_{-10.9}$ | $25.7^{+3.4}_{-3.4}$ | $0.0^{+13.7}_{-13.4}$ | $24.7^{+3.5}_{-3.5}$ |
| 53A$\otimes$90A | $28.4^{+8.1}_{-7.7}$ | $30.9^{+2.7}_{-2.9}$ | $27.7^{+9.2}_{-8.6}$ | $30.3^{+2.8}_{-2.8}$ |
| 53B$\otimes$90B | $32.3^{+7.0}_{-6.7}$ | $29.9^{+2.3}_{-2.4}$ | $30.2^{+8.2}_{-7.6}$ | $27.7^{+2.4}_{-2.5}$ |
| 53A$\otimes$90B | $45.2^{+6.1}_{-5.9}$ | $32.4^{+2.2}_{-2.2}$ | $44.0^{+6.9}_{-6.6}$ | $30.9^{+2.2}_{-2.3}$ |
| 53B$\otimes$90A | $34.6^{+9.4}_{-8.7}$ | $31.7^{+2.9}_{-3.0}$ | $33.7^{+10.8}_{-10.0}$ | $30.8^{+3.0}_{-3.0}$ |
| 53(A+B)$\otimes$90A | $31.6^{+6.3}_{-6.0}$ | $31.3^{+2.5}_{-2.6}$ | $30.8^{+7.1}_{-6.6}$ | $30.6^{+2.5}_{-2.5}$ |
| 53(A+B)$\otimes$90B | $39.3^{+4.8}_{-4.7}$ | $31.2^{+1.9}_{-2.0}$ | $37.7^{+5.3}_{-5.1}$ | $29.3^{+2.0}_{-2.0}$ |
| 53A$\otimes$90(A+B) | $37.7^{+5.1}_{-4.9}$ | $31.7^{+1.9}_{-1.9}$ | $36.8^{+5.7}_{-5.5}$ | $30.6^{+2.0}_{-2.0}$ |
| 53B$\otimes$90(A+B) | $33.5^{+5.8}_{-5.7}$ | $30.8^{+2.1}_{-2.1}$ | $31.9^{+6.7}_{-6.3}$ | $29.3^{+2.1}_{-2.1}$ |
| 53(A+B)$\otimes$90(A+B) | $35.7^{+4.0}_{-3.9}$ | $31.2^{+1.6}_{-1.7}$ | $34.4^{+4.4}_{-4.3}$ | $29.9^{+1.7}_{-1.7}$ |
| (53+90)A$\otimes$(53+90)B | $35.6^{+4.3}_{-4.2}$ | $30.7^{+1.7}_{-1.7}$ | $34.4^{+4.9}_{-4.6}$ | $29.4^{+1.7}_{-1.7}$ |
| (53A+90B)$\otimes$(53B+90A) | $30.3^{+4.3}_{-4.3}$ | $29.9^{+1.7}_{-1.7}$ | $28.8^{+4.8}_{-4.6}$ | $28.5^{+1.7}_{-1.7}$ |

# MODEL NORMALIZATIONS

The normalization of other models of cosmological anisotropy has proceeded by a detailed reworking of the above. We report the inferred $Q_{rms-PS}$ values for a number of flat geometry, critical density, scale-invariant CDM models.

An exact calculation for the power spectrum of CMB anisotropy in this context renders an effective spectral slope (in the sense of a power law approximation, $P(k) \propto k^{n_{eff}}$, used to generate the multipole coefficients $a_\ell^2$ solely through the Sachs-Wolfe effect) slightly steeper than $n = 1$ over the angular scales probed by the $COBE$-DMR instrument ($\ell \lesssim 15$). A value in the range 1.1 - 1.15 would be most appropriate. The steepening is caused by three effects:
1) the anisotropy of the photon distribution function at the moment of decoupling.
2) the high-redshift integrated Sachs-Wolfe effect generated by the change in growth rate of matter perturbations at the post-recombination epoch due to the non-negligible contribution of photons to the total energy density.
3) the contribution of the first adiabatic/Doppler peak to the low $\ell$ multipoles.

Table 2 summarizes the normalization amplitudes inferred from the $\otimes_{RMS}$ values derived from the 53 and 90 GHz (A+B) sky maps, excluding the quadrupole. The results obtained when the quadrupole is included are in excellent agreement with these once the bias introduced by the low observed quadrupole is accounted for (see Banday et al. 1994). The steeper power law approximation fits render lower amplitudes than the exactly computed CDM power spectra, and should not be considered as sufficiently good descriptors of this class of cosmological model (see Górski, Stompor & Banday 1995). The $Q_{rms-PS}$ normalization for CDM models is also $\sim 19$ $\mu$K.

Table 2: $Q_{rms-PS}$ normalizations for specific cosmological models.

| Model | $Q_{rms-PS}$ ($\mu$K) |
|---|---|
| power law, $n = 1$ | $19.4^{+2.3}_{-2.1}$ |
| power law, $n = 1.1$ | $18.2^{+2.1}_{-1.8}$ |
| power law, $n = 1.15$ | $17.9^{+2.0}_{-1.8}$ |
| CDM ($h = 0.5$, $\Omega_b = 0.01$) | $18.8^{+2.3}_{-1.9}$ |
| CDM ($h = 0.5$, $\Omega_b = 0.03$) | $18.9^{+2.2}_{-1.9}$ |
| CDM ($h = 0.5$, $\Omega_b = 0.05$) | $18.8^{+2.2}_{-1.9}$ |
| CDM ($h = 0.5$, $\Omega_b = 0.10$) | $18.7^{+2.3}_{-1.9}$ |
| CDM ($h = 0.8$, $\Omega_b = 0.03$) | $18.8^{+2.2}_{-2.0}$ |

*Acknowledegments:* We acknowledge the efforts of those contributing to the $COBE$-DMR. $COBE$ is supported by the Office of Space Sciences of NASA headquarters. We thank Radek Stompor for providing us with CDM anisotropy power spectrum coefficients.